# Stripe order and spin dynamics in triangular-lattice antiferromagnet KErSe$_2$: A single-crystal study with a theoretical description

Gaofeng Ding,[1,2,‡] Hongliang Wo,[1,2,‡] Rui Leonard Luo,[3,‡] Yimeng Gu,[1,2] Yiqing Gu,[1,2] Robert Bewley,[4] Gang Chen[3,*] and Jun Zhao[1,2,5,6,†]

[1]State Key Laboratory of Surface Physics and Department of Physics, Fudan University, Shanghai 200433, China

[2]Shanghai Qi Zhi Institute, Shanghai 200232, China

[3]Department of Physics and HKU-UCAS Joint Institute for Theoretical and Computational Physics at Hong Kong, The University of Hong Kong, Hong Kong, China

[4]ISIS Facility, Rutherford Appleton Laboratory, STFC, Chilton, Didcot, Oxon OX11 0QX, United Kingdom

[5]Institute of Nanoelectronics and Quantum Computing, Fudan University, Shanghai 200433, China

[6]Shanghai Research Center for Quantum Sciences, Shanghai 201315, China

Abstract

The rare-earth triangular-lattice chalcogenide is a great platform for exploring both spin liquids and novel magnetic orders with anisotropic spin interactions and magnetic frustrations. Here, we report the thermodynamic and neutron scattering measurements of rare-earth triangular-lattice chalcogenide KErSe$_2$, using single-crystal samples. Our experiments revealed a long-range stripe order below 0.2 K. Although the magnetic order was three-dimensional, magnetic excitations exhibited negligible modulation along the $z$ direction, indicating very weak interlayer coupling. Furthermore, magnetic excitation developed a well-defined spin-wave dispersion with a gap of ~0.03 meV at M points. Both the stripe order and spin-wave excitations could be quantitatively understood from the anisotropic spin interactions of the Er$^{3+}$ Kramers doublets.

Introduction

A triangular-lattice antiferromagnet (TLAF) is a canonical geometrically frustrated magnet [1-4]. Rare-earth triangular-lattice magnets have recently attracted considerable interest, and they have been proposed to hold various exotic quantum magnetic states, including quantum spin liquids (QSL), intrinsic quantum Ising magnets, and multipolar hidden orders due to the combination of the geometric frustration, anisotropic exchange interaction, and magnetic multipole degree of freedom induced by spin-orbit coupling (SOC) and crystal field interaction [5-16]. It has been shown that the magnetic ground states of rare-earth triangular-lattice magnets sensitively depend on rare-earth ions and their surrounding anions. For example, although YbMgGaO$_4$ shows no signs of long-range magnetic order and supports the gapless continuum in magnetic excitations, which is the hallmark of QSL states [7,8,17], the isostructural TmMgGaO$_4$ exhibits well-defined spin-wave-like excitations that can be effectively explained by the intrinsic quantum Ising model with intertwined dipolar and multipolar orders [13,14]. Similar to YbMgGaO$_4$, NaYbCh$_2$ (Ch = O, S, Se) [9-11,18-25] was also discovered to show gapless continuous magnetic excitations and was recognized as the QSL candidate. Conversely, the Ce-based counterpart ACeCh$_2$ (A = alkali metal) compounds exhibit diverse magnetic properties. Although KCeO$_2$ and KCeS$_2$ both develop long-range magnetic orders at low temperatures, as suggested by the specific heat measurements and spin-wave-like magnetic excitations [26-29], RbCeO$_2$, on the contrary, enters a quantum-disordered ground state [30].

It would be interesting to extend the research to other rare-earth elements besides the aforementioned compounds [31-37]. It has been proposed that the Er$^{3+}$ ions host similar Kramers doublet with an effective $S_{eff}$ = 1/2 under the crystalline electric field. The wavefunction of the ground-state Kramers doublet contains a considerable $J_z$ = 1/2 component, which could exhibit appreciable quantum effects, similar to the Yb$^{3+}$ case [33,34]. Although the previous thermodynamic measurements on NaErSe$_2$ and KErSe$_2$ indicated no long-range magnetic order above 0.5 K [36], recent neutron diffraction measurements on polycrystalline KErSe$_2$ suggested a long-range stripe-type magnetic order below $T \sim 0.2$ K [37]. To unambiguously determine the magnetic structure and to unveil the underlying magnetic interactions that drive the magnetic order, detailed neutron scattering measurements on single-crystalline samples are needed.

Experiment

KErSe$_2$ single crystals were synthesized via the self-flux method. First, we synthesized the powder sample via a solid-state reaction. The high-purity K ingot, Er powder, and Se chunks were mixed in stoichiometric molar quantities K:Er:Se = 1.05:1:2 in an argon-filled glove box (a small excess amount of K was used to compensate for volatility). Then, the mixture was loaded into the quartz ampoule and sealed under vacuum. The quartz ampoule was slowly heated to 550°C, followed by soaking for 48 h. Temperature was further raised to 920°C and held for 48 h. After cooling to room temperature, the phase-pure polycrystalline KErSe$_2$ sample was ground and stored in the glove box. For single-crystal growth, the KErSe$_2$ powder precursor was mixed with K$_2$Se$_3$ flux in a mass ratio of 10:1 and sealed in the quartz tube under vacuum. Next, the quartz tube was heated to 1050°C, held for 5 h, and cooled down to 650°C at a rate of 0.8°C/h. Once back to room temperature, the excess flux was removed using water and plate-like single crystals with a mass of ~20 mg each were obtained.

The X-ray diffraction (XRD) measurements of the KErSe$_2$ single crystals were performed on a Bruker D8 Discover diffractometer with Cu $K_\alpha$ radiation. Magnetic susceptibility was measured using a magnetic property measurement system (Quantum Design), and the low-temperature heat capacity down to 0.1 K was measured in a physical property measurement system (Quantum Design) with a dilution refrigerator (DR) insert. Neutron scattering experiments were conducted on the time-of-flight cold neutron multichopper spectrometer (LET) at the Rutherford Appleton Laboratory. For the neutron scattering measurement, 2.7 g of KErSe$_2$ single crystals was coaligned in the (*H*, *K*, 0) plane. The sample was then mounted in a DR to reach a base temperature of 20 mK.

Results and Discussion

Fig. 1(a) depicts the crystal structure of KErSe$_2$. The magnetic Er$^{3+}$ ions in the ErSe$_6$ octahedral environment are arranged in a perfect two-dimensional (2D) triangular-lattice configuration in the *ab* plane. The intralayer nearest-neighbor distance between Er$^{3+}$ ions (4.147 Å) is much smaller than the interlayer separation between nearby Er$^{3+}$ triangular layers (7.588 Å) [36], which indicates much weaker interlayer interactions compared to the dominant magnetic coupling between nearest-neighbor Er$^{3+}$ moments within the *ab* plane. A series of pronounced (0 0 *l*) reflections can be indexed in the single-crystal XRD measurements, as shown in Fig. 1(b), indicating the good crystallization quality and pure phase of our single-crystal samples.

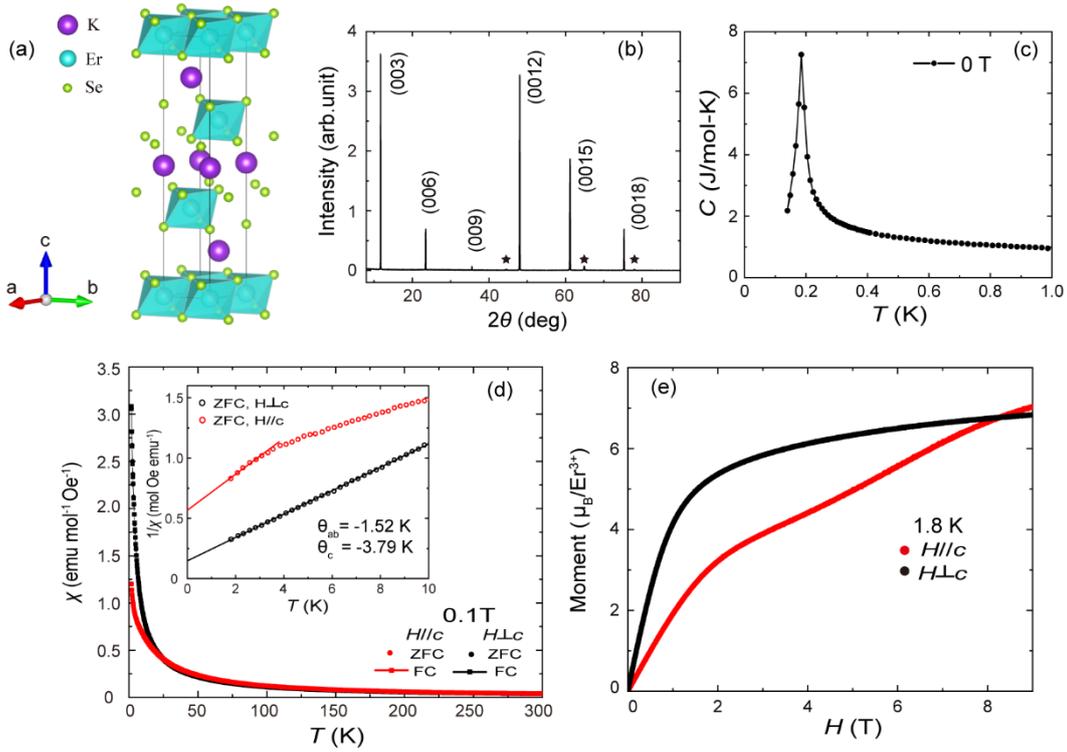

Fig. 1. (a) The crystallographic structure of KErSe$_2$. The edge-sharing ErSe$_6$ octahedrons form the 2D triangular lattice of Er$^{3+}$ ions, stacking along the $c$-axis and separated by potassium layers. (b) The (0 0 $l$) Bragg peaks in single-crystalline KErSe$_2$. The asterisks indicate the background signals from the sample stage. (c) Temperature dependence of heat capacity from 0.1 to 1 K measured under a zero field for KErSe$_2$ single crystals. (d) Field-cooling (FC) and zero-field-cooling (ZFC) measurements of magnetic susceptibility under an external magnetic field of 0.1 T applied parallel and perpendicular to the $c$-axis of KErSe$_2$. The inset shows the Curie–Weiss fit of inversed magnetic susceptibility below 10 K. (e) Isothermal magnetization for $H \parallel c$ and $H \perp c$ at $T$ = 1.8 K.

The heat capacity measurement of KErSe$_2$ single crystals down to 0.1 K is shown in Fig. 1(c). A sharp anomaly at $T \sim 0.2$ K confirms the existence of the long-range magnetic order, the nature of which was further revealed by our single-crystal elastic neutron scattering measurements (discussed below). Fig. 1(d) shows the magnetic susceptibility of KErSe$_2$ under external fields applied in the $ab$ plane and along the $c$ direction. Magnetic anisotropy appears below 25 K. The Curie–Weiss fitting in the low-temperature range yields Curie–Weiss temperatures of $\theta_{ab} = -1.52$ K and $\theta_c = -3.79$ K [inset in Fig. 1(d)]. The negative Curie–Weiss temperature indicates dominant antiferromagnetic interactions between Er$^{3+}$ ions. The $M$–$H$ curves were measured up to 9 T for both $H \parallel c$ and $H \perp c$ at $T$ = 1.8 K, as illustrated in Fig. 1(e). Although no signs of saturation were observed for both orientations in the current field range, the large anisotropy in magnetization was clearly demonstrated. The thermodynamic characterizations presented above are consistent with the previous results in Ref. [36]. The absence of saturation in the magnetization curve is attributed to the relatively small crystal field gap (approximately 0.9 meV) in KErSe$_2$ [34]. Before the polarization of the ground-state doublet under a magnetic field, the excited doublets would get involved and would significantly contribute to magnetization. This is why a 9 T magnetic field is insufficient to fully polarize the Er$^{3+}$ magnetic moments, although the exchange energy scale (from the Curie–Weiss temperature) for the lowest doublets is relatively small.

Moreover, a weak crystal field gap should generate considerable Van Vleck susceptibility, which would complicate the extrapolation of the low-temperature Curie–Weiss law. In fact, we expect the actual Curie–Weiss temperatures after subtracting the Van Vleck susceptibility to be much reduced compared to the current fitted values. Therefore, we made an attempt in the supplementary material and obtained reduced Curie–Weiss temperatures.

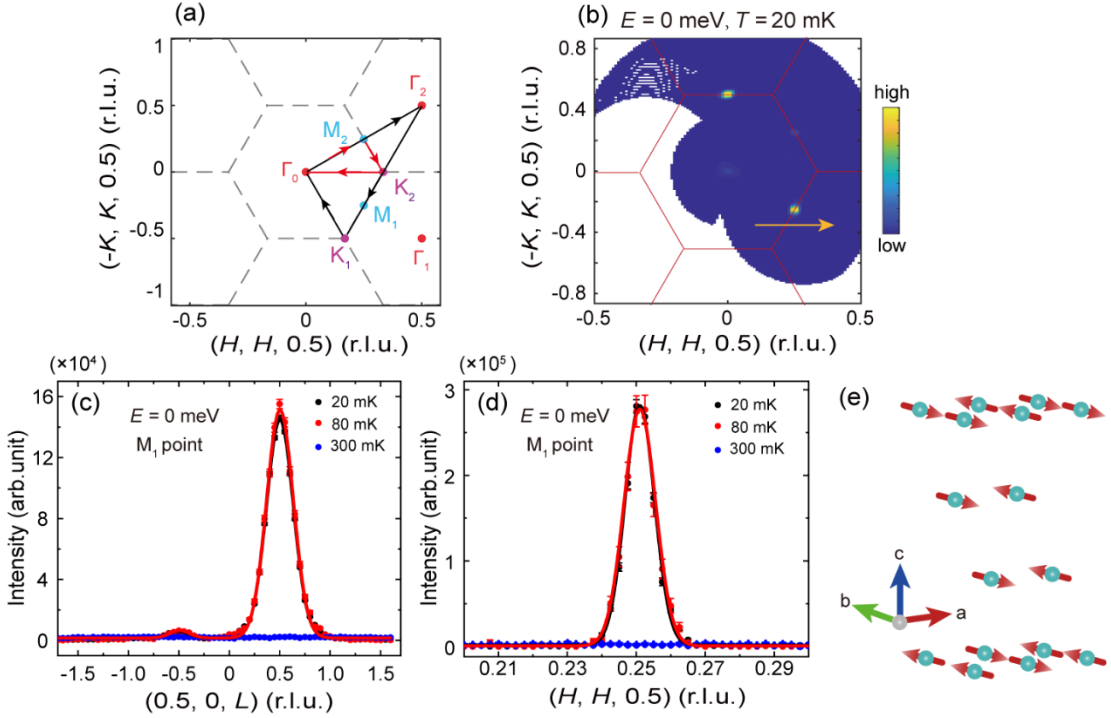

Fig. 2. (a) Sketch of the reciprocal lattice of KErSe$_2$. The Γ, M, and K points are in red, cyan, and purple, respectively. (b) The elastic neutron scattering signals in the $(H, K, 0.5)$ plane at $T = 20$ mK. The red line indicates the Brillouin zone boundaries. The yellow arrow marks the direction of $Q$-cuts across the M$_1$ point in (d). (c) $L$-cuts and (d) $(H, H)$-cuts for the magnetic Bragg peaks at the indicated temperatures. The solid lines are the fitting results of the Gaussian profile. The weak peak at $L = -0.5$ in (c) originates from the misalignment of minor single crystals. (e) Magnetic structure of KErSe$_2$ with $k = (1/2, 0, 1/2)$. Only half of the doubled magnetic unit cell along the $c$-axis is shown for clarity. The magnetic moments lying in the $ab$ plane are represented by red arrows.

The single-crystal elastic neutron scattering measurements in Fig. 2 provide a deeper insight into the magnetic structure of KErSe$_2$. Fig. 2(a) displays the hexagonal reciprocal lattice of KErSe$_2$, in which several high-symmetry points are highlighted. Clear magnetic Bragg peaks at the M points are revealed in the contour plot of elastic scattering in the $(H, K, 0.5)$ plane at the lowest temperature [Fig. 2(b)], which is further confirmed by the $Q$-cuts in Fig. 2(d). Additionally, the cuts along $(0.5, 0, L)$ [Fig. 2(c)] show a sharp peak centered at $L = 0.5$, indicating the three-dimensional nature of the magnetic order. The magnetic Bragg peaks disappear on warming to 300 mK, which is consistent with the phase transition in the heat capacity measurement. A stripe-type magnetic structure with the propagation wave vector $k = (1/2, 0, 1/2)$, in which the magnetic moments were aligned parallel (antiparallel) along the $b$ ($a$) axis, was further determined by our Rietveld refinement, as shown in

Fig. 2(e) (for the details of the refinement, see supplementary material). This stripe order is precisely one of the stripe orders predicted with the anisotropic spin model for the spin-orbit-coupled TLAF away from the XXZ limit [5].

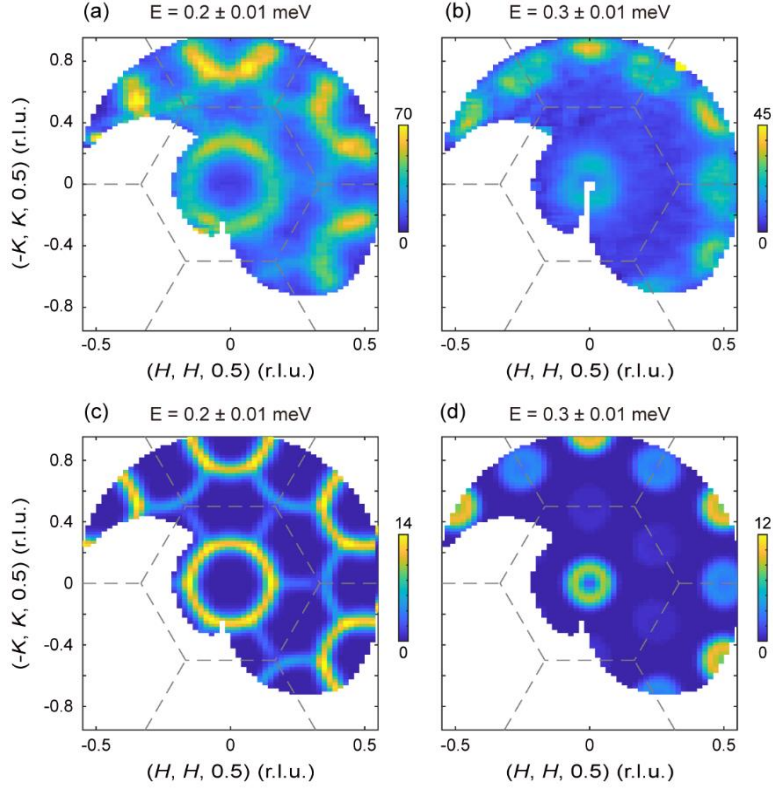

Fig. 3. The measured and calculated momentum dependence of spin excitations in $KErSe_2$ at indicated energies. (a, b) ($H$, $K$, 0.5) contour plots with energy transfer $E = 0.2$ and 0.3 meV measured at $T = 20$ mK. (c, d) Corresponding calculated spin excitations using the model specified in the text. The dashed lines indicate the zone boundaries. The color bars indicate the scattering intensity of arbitrary units on a linear scale.

    Now, we turn to the spin excitations in $KErSe_2$, which reveal the magnetic interactions between the local moments of the $Er^{3+}$ ions. Fig. 3 illustrates the in-plane momentum dependence of spin excitations at 20 mK. Sharp spin-wave-like excitations with sixfold symmetry can be observed, as expected for the underlying long-range magnetic order. The detailed excitation behavior is more clearly demonstrated in Fig. 4(a) and 4(b), which shows the momentum–energy dispersions along the high-symmetry directions. The spectrum shows a ~0.03 meV spin gap at the M points [Fig. 4(g)], which is a natural result due to the absence of continuous symmetry in the anisotropic spin model [5] and which will be discussed later.

    Interestingly, the spin gap at the K point is only slightly larger than that at the M point, as shown in Fig. 4(f, g), indicating the proximity of the magnetic ground state to the phase boundary between the stripe and 120° order. With increasing energy transfers, the sharp spin-wave excitations disperse outward and finally reach the energy zone boundary below 0.4 meV, forming a ring-like pattern around the Γ point. Furthermore, the featureless spin excitations along the $L$ direction [Fig. 4(e)] suggest negligible interlayer magnetic interactions.

To understand the nature of the magnetic order and spin excitations of KErSe$_2$, we attempted to develop the microscopic model of the exchange interactions. The isolated Er$^{3+}$ ion has a 4$f^{11}$ electronic configuration with $L = 6$, $S = 3/2$. The strong SOC in 4$f$-electron compounds results in an effective $J = 15/2$ with 16-fold degeneracy for KErSe$_2$. Just like the Yb$^{3+}$ ions in triangular-lattice YbMgGaO$_4$ [7,8] and AYbCh$_2$ (A = Na, K, Cs; Ch = O, S, Se, Te) [9-12,18,19,21-23], the Er$^{3+}$ ions in KErSe$_2$ are located in the same $D_{3d}$ crystal field environment. The ground state can therefore be expected to be a Kramers doublet. Moreover, the usual Kramers doublet ground state [15] with a large $|J_z = \pm 1/2 >$ component has been confirmed in KErSe$_2$ by crystal field analysis in Ref. [34]. As a result, we naturally expect the anisotropic spin Hamiltonian of KErSe$_2$ to share the same following form as the one in YbMgGaO$_4$ [5,17].

$$H = \sum_{\langle ij \rangle} J_{zz} S_i^z S_j^z + J_{\pm}(S_i^+ S_j^- + S_i^- S_j^+) + J_{\pm\pm}(\gamma_{ij} S_i^+ S_j^+ + \gamma_{ij}^* S_i^- S_j^-)$$
$$- \frac{iJ_{z\pm}}{2}[(\gamma_{ij}^* S_i^+ - \gamma_{ij} S_i^-)S_j^z + S_i^z(\gamma_{ij}^* S_j^+ - \gamma_{ij} S_j^-)], \quad (1)$$

where the coupling $J_{\pm\pm}$ and $J_{z\pm}$ terms represent the bond-dependent anisotropic interactions arising from the strong SOC, $S_i^\alpha$ ($\alpha = x, y, z$) indicates the spin–1/2 operators acting on the doublet at site $i$, $S_i^\pm = S_i^x \pm iS_i^y$ are the ordinary ladder operators, $\gamma_{ij}$ represents the phase factor defined in Ref. [5,6], and $\langle ij \rangle$ is the sum over nearest Er$^{3+}$ neighbors. It is clear that in the absence of the anisotropic $J_{\pm\pm}$ and $J_{z\pm}$ terms, the Hamiltonian reduces to an XXZ model with only $J_{zz}$ and $J_\pm$ coupling. The XXZ limit would mainly favor a magnetic order with the ordering wavevector at the K points, and thus, it is not relevant for the stripe order here.

On the basis of the abovementioned effective Hamiltonian, we used the SpinW program [38] to perform the quantitative linear spin-wave theory calculations of the spin excitations for representative sets of exchange parameters. Herein, we neglected both next-nearest-neighbor (NNN) and interlayer couplings in our calculation due to the strong localization of the 4$f$ electron wavefunction and the quasi-2D spin excitations [Fig. 4(e)]. The calculated ($H$, $K$, 0.5) constant-energy plots and the *Q–E* dispersions were compared to the experimental data in Fig. 3 and Fig. 4, both of which showed excellent agreement. The overall shapes of the spin excitations, as well as the experimentally observed energy gaps, were accurately captured by the calculated excitation spectrum. The minor intensity discrepancy between the data and calculations could be attributed to the NNN interaction terms and potential quantum fluctuations not included in our model and the linear spin-wave theory.

The consistency between our simulations and experimental data allows the quantitative determination of the exchange parameters in KErSe$_2$. Our numerical simulations discovered a representative set of exchange parameters with $J_{zz} = 0.06$ meV, $J_\pm = 0.01$ meV, $J_{\pm\pm} = -0.04$ meV, and $J_{z\pm} = 0.06$ meV. Note that our simulation results are relatively insensitive to the $J_{z\pm}$ term, which can vary in a certain range of 0.01–0.06 meV (−0.01 to −0.06 meV). The two anisotropic interaction terms $J_{\pm\pm}$ and $J_{z\pm}$ both have non-negligible values comparable to $J_{zz}$. Therefore, not only the bond-independent interactions ($J_{zz}, J_\pm$) but also the bond-dependent anisotropic terms ($J_{\pm\pm}, J_{z\pm}$) would be the key ingredients for the stripe order and spin dynamics in KErSe$_2$. Theoretical studies on the phase diagram of the anisotropic spin–1/2 TLAF model suggested a large region of bond stripe order with prominent $J_{\pm\pm}$ and $J_{z\pm}$ interactions [5,6], in which KErSe$_2$ is expected to be located. Because the Curie–Weiss temperatures are only related to $J_{zz}$ and $J_\pm$ [5], it seems that the fitted $J_{zz}$ and $J_\pm$ from the spin-wave dispersion are relatively small compared to the extracted Curie–Weiss temperatures. As previously stated, without careful subtraction, the extracted Curie–Weiss temperatures would be higher than the actual Curie–Weiss temperatures that govern the low-temperature magnetic properties of the ground-state doublets.

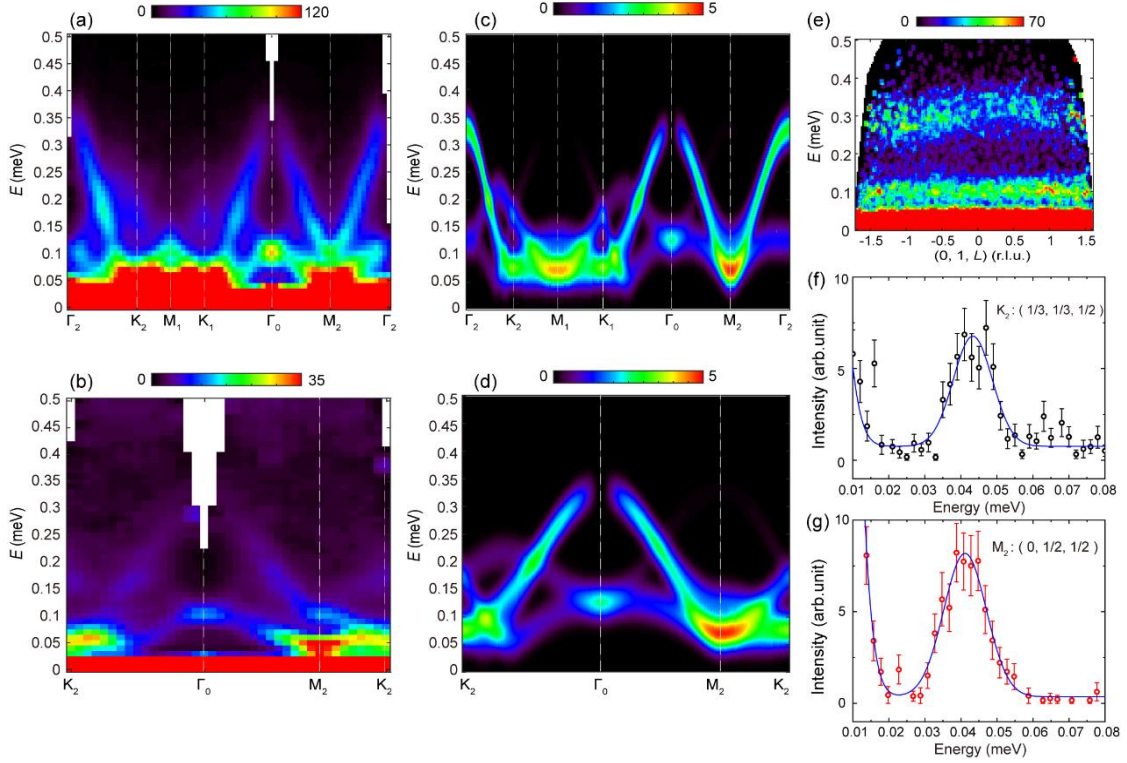

Fig. 4. The measured and calculated spin-wave dispersions in KErSe$_2$. (a) and (b) show the dispersive behaviors of the spin excitations along the two high-symmetry directions marked by the black and red arrows in Fig. 2a. Data were collected at $T = 20$ mK. (c, d) The corresponding calculated spin excitation dispersions using the model mentioned in the main text. The color bars indicate the scattering intensity of arbitrary units on a linear scale. (e) The $L$-dependence of spin excitations at $\Gamma_2$ point. (f, g) Constant $Q$-cuts at K$_2$ and M$_2$ points demonstrating the energy gap. The blue solid line shows the Gaussian fitting of the data. The data in (a) and (e) were measured with incident energy $E_i = 1.8$ meV, and those in (b), (f), and (g) were measured with $E_i = 0.8$ meV.

It is also instructive to compare the magnetic ground states of triangular-lattice KErSe$_2$ with those of its rare-earth-based cousins. For the Yb$^{3+}$-based AYbCh$_2$ with $S_{\text{eff}} = 1/2$, the absence of a long-range magnetic order, together with the observation of a gapless continuum by the inelastic neutron scattering measurements, indicated the QSL ground state [9-12,18,19,21,23]. Conversely, though sharing the same $S_{\text{eff}} = 1/2$ ground state, Ce$^{3+}$-based analog compounds KCeO$_2$ and KCeS$_2$ developed long-range magnetic orders at low temperatures, as suggested by specific heat measurements and spin-wave-like magnetic excitations [26-28]. In our current research on KErSe$_2$, the stripe magnetic order and spin-wave-like magnetic excitations have been revealed, which indicate a more classical behavior than Yb$^{3+}$ species. The key could lie in the variety of the single-ion ground states. Although the ground-state doublet in KErSe$_2$ possesses a large $|\pm 1/2\rangle$ component enduring the potential for appreciable quantum effects [34], the concomitant substantial $|\pm 7/2\rangle$ and $|\pm 13/2\rangle$ components would strongly suppress the quantum effects. At the same time, anisotropic interaction $J_{\pm\pm}$ favors the magnetic order state [5,6]. Therefore, the large ratio of $J_{\pm\pm}$ and $J_{zz}$ interactions ($|J_{\pm\pm}/J_{zz}| = 2/3$) in KErSe$_2$, which is much more prominent than the one in YbMgGaO$_4$ ($|J_{\pm\pm}/J_{zz}| \simeq 0.1$) [17], could explain the less frustrated magnetic ground state. Notably, similar to KErSe$_2$, KCeS$_2$ also develops the stripe magnetic order but with a different stripe-yz type [29]. According to the generic phase diagram of Ref [5], KErSe$_2$ and KCeS$_2$ are located at two distinct stripe-order phases with different spin orientations, as evidenced by their spin interaction [29].

Conclusion

In summary, we observed a long-range stripe-type magnetic order with the propagation wave vector ***k*** = (1/2, 0, 1/2) in the rare-earth triangular-lattice antiferromagnet KErSe$_2$ below $T_N$ = 0.2 K. Inelastic neutron scattering measurements of KErSe$_2$ single crystals were performed to study the low-energy spin excitations, which reveal the sharp spin-wave-like modes. Moreover, the spin excitation spectra can be effectively described by an anisotropic Heisenberg model with nearest-neighbor exchange interactions. Finally, on the basis of the great consistency between our theoretical treatment and experimental data, spin–spin exchange interaction parameters of KErSe$_2$ have been quantitatively extracted.

Acknowledgments

We would like to thank Dr. Bowen Ma for the helpful discussion. This work was supported by the Innovation Program of the Shanghai Municipal Education Commission (Grant No. 2017-01-07-00-07-E00018), the National Natural Science Foundation of China (Grant No. 11874119), and the Shanghai Municipal Science and Technology Major Project (Grant No. 2019SHZDZX01). G.C. was supported by the Research Grants Council of Hong Kong with General Research Fund Grant No. 17306520. H.W. acknowledges support from the China National Postdoctoral Program for Innovative Talents (Grant No. BX2021080), China Postdoctoral Science Foundation (Grant No. 2020M700860), and Shanghai Post-doctoral Excellence Program (Grant No. 2021481).

[†]zhaoj@fudan.edu.cn
[*]gangchen@hku.hk
[‡]These authors contribute equally to this work.